# SOT-MRAM Bitcell Scaling with BEOL Read Selectors: A DTCO Study

Y. Xiang, F. García-Redondo, A. Sharma, V. D. Nguyen, A. Fantini, P. Matagne, S. Rao,
S. Subhechha, L. Verschueren, Md. A. Baig, M. Garcia Bardon, G. Hellings

*Abstract*—This work explores the cross-node scaling potential of SOT-MRAM for last-level caches (LLCs) under heterogeneous system scaling paradigm. We perform extensive Design-Technology Co-Optimization (DTCO) exercises to evaluate the bitcell footprint for different cell configurations at a representative 7 nm technology and to assess their implications on read and write power-performance. We crucially identify the MTJ routing struggle in conventional two-transistor one-resistor (2T1R) SOT-MRAMs as the primary bitcell area scaling challenge and propose to use BEOL read selectors (BEOL RSs) that enable (10 – 40) % bitcell area reduction and eventually match sub-N3 SRAM. On writability, we affirm that BEOL RS-based bitcells could meet the required SOT switching current, provided the magnetic free layer properties be engineered in line with LLC-specific, (0.1 – 100) s retention targets. This is particularly to attribute to their (i) more available Si fins for write transistor and (ii) lower bitline resistance at reduced cell width. We nevertheless underscore the read tradeoff associated with BEOL RSs, with the low-drive IGZO-FET selector sacrificing the latency up to (3 – 5) ns and the imperfectly rectifying diode selectors suffering (2.5 – 5) × energy cost relative to 2T1R. This article thus highlights the realistic prospects and hurdles of BEOL RSs towards holistic power-performance-area scaling of SOT-MRAM.

## I. Introduction

THE RECENT age has been witnessing a skyrocketing in the demand for high-capacity on-chip memories, driven largely by high-performance computing (HPC) and AI accelerators [1][2]. Indeed, in the epoch where hundreds of MBs of last-level caches become essential to compute scaling [1][2], traditional SRAM faced with capacity and leakage challenges starts to struggle [3]. 3D-bonded SRAMs have been introduced to address the capacity issue [4], while for the desired low leakage one often has to look to non-volatile memories to the rescue [2], among which the high-endurance and sub-ns-writing SOT-MRAM stands out [5]. Notwithstanding the spotlight, the area overhead and high write current associated with SOT-MRAM have long been a major concern [2]. Multi-bit, voltage gating-assisted SOTs (VGSOTs) have in response been proposed to salvage both the density-per-bit and write current [6][7], but the required voltage-controlled magnetic anisotropy has so far yet to overcome prohibitive technological challenges [7].

The latest rise of CMOS 2.0 paradigm [8], on the other hand, seems to be opening up new opportunities to SOT-MRAM for LLC scaling. In this new paradigm where the envisioned smart system function partitioning would allow the "slow" LLCs to move away from the dense logic and/or level-1/-2 cache (L1/L2) tiers, one may, instead of attempting to *outscale and replace* the fast and densest SRAMs made of most advanced technologies such as CFETs [9], pivot to *complementing* SRAM with *cross-node matched* density and hence capacity of SOT-MRAMs made of relaxed, cost-effective technologies in a heterogeneous fashion (Fig. 1(a)). Indeed, as we will show in the paper, such opportunities crucially hinge on the enablement of BEOL read selectors. Moreover, for write current scaling, we shall point out that it could be reduced to an affordable level in scaled, logic-transistor bitcells if the magnetic properties of the MTJ were made aligned to LLC-relevant retention targets [1][3], and this would in fact be facilitated by BEOL read selectors.

The paper proceeds as follows: in Sec. II we outline the footprint scaling target for *cross-node matching*, which is followed by the analyses of 2T1R bitcell bottlenecks and the proposal of compact BEOL read selector (both three- and two-terminal)-based bitcells in Sec. III. Then we discuss the write current, retention and speed associated with the BEOL selector-based bitcells in Sec. IV, after which the read design space and power-performance tradeoff are analyzed in Sec. V. Sec. VI summarizes the main learnings of the work.

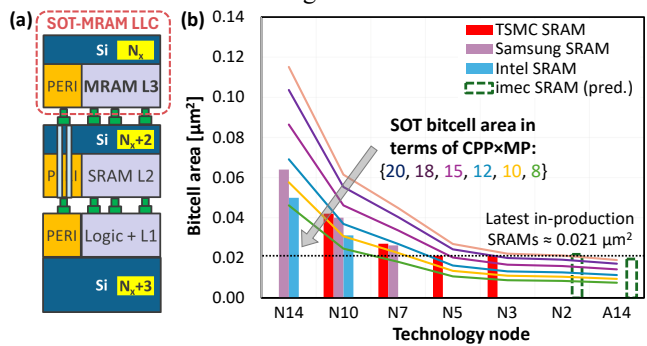

Fig. 1. (a) Concept of cross-node density matching of SOT-MRAM as LLCs (e.g., L3) using a relaxed "$N_x$" node that is hybrid-bonded to L1/L2

The authors are with imec, Leuven, 3001 Belgium.
This work has been enabled in part by the NanoIC pilot line. The acquisition and operation are jointly funded by the Chips Joint Undertaking, through the European Union's Digital Europe (101183266) and Horizon Europe programs (101183277), as well as by the participating states Belgium (Flanders), France, Germany, Finland, Ireland and Romania. For more information, visit nanoic-project.eu.
This work has been additionally supported by Imec's Industry Affiliation Program (IIAP).

SRAMs in more advanced nodes yet at comparable density. (b) Top-down profiling of SOT-MRAM bitcell area in terms of how many CPP×MP needed to achieve cross-node matching with respect to SRAM scaling roadmap [10].

## II. Scaling Target for Cross-Node Density Matching

As mentioned above, the *cross-node matching* of SOT-MRAM despite no longer pursuing supra-SRAM density still holds iso-SRAM density as essential such that one may anticipate no LLC capacity loss (Fig. 1(a)). This accordingly necessitates that the bitcell area of the SOT-MRAM be no greater than latest in-production SRAMs, which today sits at 0.021 μm² in N3 [11]. Knowing that SOT-MRAM bitcell dimensions in (post-)FinFET era can be broadly estimated in terms of the number of contact poly pitches (CPPs) for cell width and of the number of the minimum metal pitches (MPs) for cell height [12], here in Fig. 1(b) we perform a top-down profiling of SOT-MRAM bitcell area in units of CPP×MP. Evidently, the earliest node we could revert to would be the 7 nm node that still intercepts the 0.021 μm² benchmark, when the cell area is below 8×CPP×MP. In other words, for cost-effective *cross-node matching*, the least costly option would be to scale SOT-MRAM bitcell at N7 to under 8×CPP×MP.

In the work below, we will use our inhouse estimations and hardware data for the N7 FinFET bitcell transistor (Table I) [13] and SOT-MRAM (Table II) assumptions, respectively, while the BEOL parasitics are captured by Raphael™ FX simulations [14]. Note that for electrical simulations in this work, the "*typical*" (i.e., nominal) corner of the N7 bitcell transistor will be used by default unless otherwise specified.

TABLE I. Tech. Assumptions Bitcell Transistor (N7 NMOS FinFET)

| Parameter | Description [unit] | Value (incl. process variation) |
|---|---|---|
| CPP | Contact poly pitch [nm] | 56 |
| MP | Minimum metal pitch [nm] | 40 |
| FP | Fin pitch [nm] | 30 |
| $L_g$ | Drawn gate length [nm] | 21 ± 0.6 |
| HFIN | Fin height [nm] | 45 ± 1.33 |
| TFIN | Fin width [nm] | 5 ± 0.17 |
| EOT | Equivalent oxide thickness [nm] | 0.808 ± 0.037 |
| $\Phi_g$ | Gate work function [eV] | 4.326 ± 0.008 |

TABLE II. Device Parameters for SOT-MRAM

| Parameter | Description [unit] | Exp. value | Extrap. value |
|---|---|---|---|
| $D_{MTJ}$ | MTJ diameter [nm] | 63 | - |
| $\Delta$ | Thermal stability factor at 298 K [$k_BT$] | ~ 40 | 38 – 64 |
| $w_{SOT}$ | SOT track width [nm] | 99 | 73 |
| $d_{SOT}$ | SOT track thickness [nm] | 3.5 | - |
| $B_k$ | Effective anisotropy field [mT] | 219 | - |
| $B_x$ | In-plane field provided by integrated magnetic hardmask [15] [mT] | 25 | - |
| TMR | Tunneling magnetic ratio [%] | 86 | 100 – 200 |
| RA | MTJ resistance-area product [Ω·μm²] | 11 | 10 – 1000 |
| $d_{FL}$ | Free layer thickness [nm] | 1 | - |
| $\theta_{SH}$ | Spin-Hall angle | 0.158 | 0.3 |

## III. Footprint Scaling with BEOL Read Selectors

### 1) MTJ routing challenge in 2T1R bitcells

The conventional 2T1R SOT-MRAM uses a read transistor (RDT) and a write transistor (WRT) both in the FEOL active area, with a shared WL routed in poly, while the RDT drain is internally routed to the MTJ top (Fig. 2(a)). Such "internal routing" could bring unwanted complexities to integration as the MTJ top must dive down to the RDT drain pin on M2 by a total stack height of over 300 nm (Fig. 2(b)(c)). Assume a 25 nm line width for M2, direct MTJ internal routing would create a flagpole-like via (FV) with an aspect ratio (AR) greater than 10, that would pose an exorbitant integration challenge (normally via AR < 3) and render 2T1R-FV rather unrealistic.

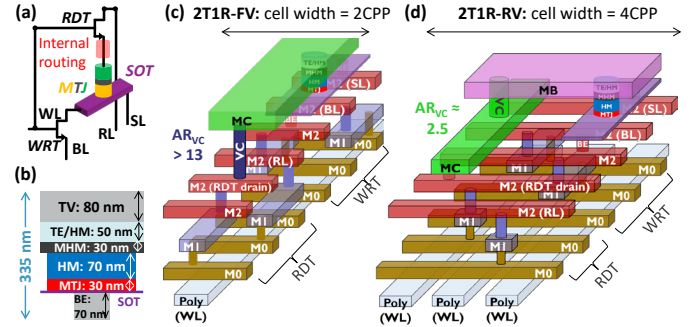

Fig. 2. (a) Internal routing node for 2T1R SOT-MRAM bitcell between the MTJ top and the RDT drain. (b) Cross-section of a typical MTJ/SOT stack corresponding to the internal routing node, with a total height of 335 nm. The stack consists of layers of bottom electrode (BE), SOT, MTJ, hardmask (HM), magnetic hardmask (MHM), top electrode as hardmask (TE/HM) and top via (TV). (c)(d): Two possible configurations for 2T1R SOT-MRAM, with (c) a flagpole-like via (FV) namely VC at AR > 13 and (d) a reduced via (RV) namely VC at AR ≈ 2.5.

An alternative to avoid the FV is to route the MTJ top first transversally and then stairwise down (Fig. 2(d)), in which case a reduced via (RV) of AR around 2.5 would be possible, but the cell width would then increase from 2×CPP in FV to 4×CPP and hence penalize bitcell area. The dilemma between via AR and cell width thus creates an impasse to 2T1R scaling that ultimately results from the "down-routing" nature of MTJ as an "inverted via" that rises above but ends on M2.

### 2) BEOL read selector-enabled density scaling

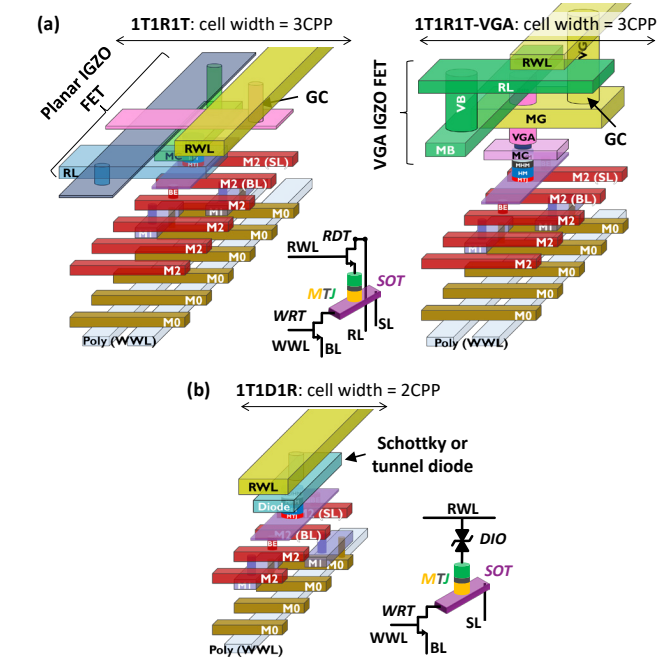

Fig. 3. Schematic drawing of (a)(b) 1T1R1T and (c) 1T1D1R bitcells. (a) 1T1R1T cells using a planar (left [16]) or vertical gate-all-around ("VGA",

right [17]) IGZO-FETs with a separate read wordline (RWL). (b) 1T1D1R cell with a tunnel- [18] or Schottky- [19][20] diode.

To fundamentally lift the internal "down-routing" limitation in SOT-MRAM, here we propose to build the read selector *on top* of the MTJ in BEOL, upon which the MTJ top could be conveniently routed *upwards* – or in other words, which would allow the MTJ to serve as a "normal via"! This is made possible by recent development of BEOL-compatible devices, either in the form of three-terminal transistors (e.g., IGZO-channel FETs [16][17]) or of two-terminal nonlinear diodes (e.g., symmetric tunnel- [18] or asymmetric Schottky-diode [19][20]). Based on this, we propose the following two SOT-MRAM bitcell configurations: one-transistor, one-resistor (MTJ), one-BEOL transistor (1T1R1T; Fig. 3(a)) with a vertical-channel gate-all-around (VGA; [17])-variant, and one-transistor, one-diode and one-resistor (1T1D1R; Fig. 3(b)).

As evidenced in Fig. 4, both 1T1R1T(-VGA) and 1T1D1R permit notable bitcell dimension shrinking, particularly in cell width. Indeed, by seating the selector directly on top of the MTJ, one can finally remove the bulky metal line MC used for down routing in 2T1R-RV that runs parallelly in the same plane with SOT (Fig. 2(d)) and roll the cell width back by at least one CPP. Since CPP is larger than MP in technology ground rules (CPP = 57 nm, MP = 40 nm), the cell width scaling enabled by using BEOL RSs turns out to be sufficient for area reduction, even if the planar IGZO-FET in 1T1R1T requires one extra MP along in cell height to accommodate its contact pitch (> 90 nm). Further reduction in bitcell width becomes less obvious in IGZO-FET-based bitcells due to the additional transversal space required for fitting in the gate contact (GC) via created by the separate read wordline (RWL; Fig. 3(a)) that needs to sit outside of the IGZO active, yet it could still be overcome by adopting self-aligned GC (SAGC) [21] and/or fully customized BEOL [22] provided processes and costs allow. The 1T1D1R bitcell, on the other hand, can successfully scale the cell width to 2×CPP thanks to the self-aligned diode formation, thus achieving sub-N3 SRAM bitcell area (Fig. 4). Overall, compared to either the 2× oversized 2T1R-RV or the high-AR 2T1R-FV, the BEOL RSs *in realistic terms* do prove to offer an unsophisticated route to SOT-MRAM *cross-node matching*.

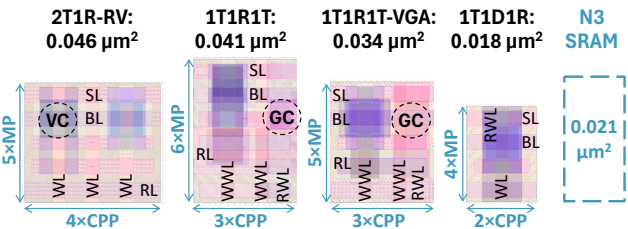

Fig. 4. SOT-MRAM bitcell layout and area of different configurations and illustrative N3 SRAM cell area (0.021 μm² [11]). Note that the (W)WLs in 2T1R-RV, 1T1R1T and 1T1R1T-VGA run in multiple fingers. The "VC" and "GC" match the descriptions in Fig. 2(d) and Fig. 3(a), respectively.

## IV. WRITABILITY ANALYSES IN LLC CONTEXT

The high write current associated with the Spin-Hall effect (SHE) has been another major skepticism about SOT-MRAM bitcell footprint scaling in literature [2], and it is indeed critical to examine whether this could fit in the *cross-node matching* paradigm proposed in this work. To do this, we have simulated in SPICE [23] the bitcell supply current ($I_{cell}$) in the farthest cell in a 2-KB array (128×128; Table III) based on the proposed bitcell configurations in Fig. 2, Fig. 3 and Fig. 4, using the N7 transistor in Table I across the typical ("*tt*"), slow ("*ss*") and fast ("*ff*") process corners. Remarkably, the most compact 1T1D1R bitcell despite having only a number of fingers (NF) of one does not suffer from any $I_{cell}$ loss compared to 2T1R-RV that is populated by three WRT fingers even in the worst-case "*ss*" corner, but on the contrary even shows a notable $I_{cell}$ increase (+ 8 %) in "*tt*" corner. This has to do first with the removal of Si RDT and the associated contacts, which increases the number of fins (NFIN) per finger, and second with the halving of BL resistance thanks to the bitcell width reduction. Given the large BL resistance along 128 rows in the array that is on par with or even eclipsing the SOT resistance (Table III), the cell width scaling is in practice an integral part to the write current scaling, which undoubtfully concurs with the density scaling in *cross-node matching*. In this sense, the wide-cell, multi-finger nature of 1T1R1T cells *per se* does not lend much favor to $I_{cell}$ but it is the removal of Si RDT that maximizes the available NFIN and gives them up to an extra 22 % $I_{cell}$ benefit ("*ss*" corner).

TABLE III. WRITE SPACE IN 2-KB (128×128) ARRAY (@ $V_{WRITE}$ = 0.7 V [a])

|  | 2T1R-RV | 1T1R1T | | 1T1D1R |
|---|---|---|---|---|
|  |  | Planar | VGA |  |
| $NF_{WRT}$ | 3 | 2 | 2 | 1 |
| $NFIN_{WRT}$ per finger | 3 | 6 | 5 | 4 |
| BL resistance [kΩ] | 1.7 | 1.4 | 1.4 | 0.87 |
| SOT resistance [kΩ] | 0.66 | 0.66 | 0.66 | 0.66 |
| $I_{cell}$ [μA] | 115 (*ss*) | 140 (*ss*) | 135 (*ss*) | 115 (*ss*) |
|  | 124 (*tt*) | 152 (*tt*) | 147 (*tt*) | 134 (*tt*) |
|  | 130 (*ff*) | 159 (*ff*) | 156 (*ff*) | 154 (*ff*) |

[a]: with 100 mV gate overdrive.

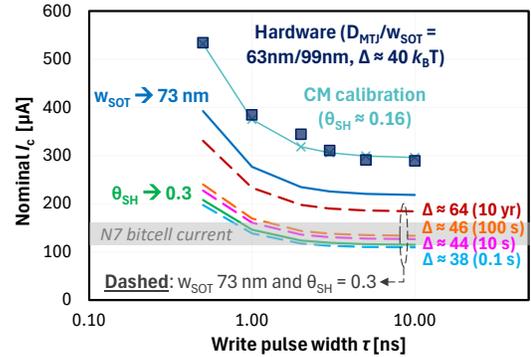

Fig. 5. Extrapolated nominal SOT $I_c$ versus write pulse width τ using the model in Eq. (1) calibrated to hardware data. The solid lines (no symbols) indicate geometry ($w_{SOT}$) and material ($θ_{SH}$) scaling extrapolation based on prior work [5] and literature [27], respectively, while the dashed curves apply additionally different retention targets calculated per [28][29] (see Table IV). Also shown is the bitcell current range computed in Table III.

To put the available $I_{cell}$ numbers in real SOT switching perspective, we have developed a time-dependent, macrospin SOT switching model (Eq. (1)) based on [24], that is calibrated on inhouse hardware (Fig. 5). Here we associate the switching current as a function of pulse width $I_c(τ)$ with not only various material (e.g., $θ_{SH}$) and geometry (e.g., $D_{MTJ}$, $w_{SOT}$) parameters but also the thermal stability factor Δ that gauges the retention:

$$I_c(\tau) = \frac{2eM_s d_{FL} w_{SOT} d_{SOT}}{\hbar \theta_{SH}} \left(\frac{B_k}{2} - \frac{B_x}{\sqrt{2}}\right) \sqrt{1 + \left(\frac{\pi \tau_D}{\tau}\right)^2}$$

$$= \frac{8 e k_B T w_{SOT} d_{SOT} \Delta}{\pi \hbar \theta_{SH} D_{MTJ}^2} \left(1 - \frac{\sqrt{2} B_x}{B_k}\right) \sqrt{1 + \left(\frac{\pi \tau_D}{\tau}\right)^2}, \quad (1)$$

with $\tau_D$ referring to the "dynamic time scale" defined in [25]; here the relation between $\Delta$ and magnetic free layer properties ($B_k$, $M_s$, $d_{FL}$, $D_{MTJ}$) is invoked per Eq. (2):

$$\Delta = \frac{1}{k_B T} \frac{B_k M_s}{2} \pi \left(\frac{D_{MTJ}}{2}\right)^2 d_{FL}. \quad (2)$$

We emphasize that the estimated switching current $I_c(\tau)$ here only considers the nominal case, as in reality the $I_c(\tau)$ is more often than not subject to both process variations [5] and more fundamentally, thermal agitation-induced stochasticity [26].

TABLE IV. MTJ Free Layer Δ to Retention Conversion

| Retention time [a] | 0.1 s | 1 s | 10 s | 100 s | 10 yrs |
|---|---|---|---|---|---|
| Δ calculated at operating temp. 353 K [28] | 32.3 | 34.6 | 36.9 | 39.2 | 54.2 |
| Δ renormalized to $k_B T$ at 298 K | 38.2 | 41.0 | 43.7 | 46.4 | 64.2 |

[a]: per error rate of $10^{-6}$ [29].

Knowing the required nominal SOT switching current per $I_c(\tau)$ and the available $I_{cell}$ range per Table III, it is apparent that both SOT line width and SHE efficiency need to be brought up to realistic state-of-the-art in order for $I_{cell}$ to intercept, that is, $w_{SOT}$ down to 10-nm margin over $D_{MTJ}$ [5] and $\theta_{SH}$ up to the β-tungsten limit of 0.3 [27] (Fig. 5). More realistically though, in terms of the Δ that defines the data retention (see Table IV), it is Δ ∈ (38, 46) namely (0.1 – 100) s retention that one ought to target given the available Si area for WRT in *cross-node matching* and the process variations thereof – instead of the technology-wise most-used [5][6] 10-year retention (Δ ≈ 64). Fortunately, such tradeoff in retention is not entirely unjustifiable in view of the common DRAM retention time (64 ms [2]), the system requirement for global caches (~ $10^0$ s [2]) and the specifications of recent industry STT-MRAMs (~ 10 s [1][3]). To this end, the BEOL RS-based bitcells at a representative 2-KB (128×128) array-level, are at least on par with (1T1D1R) or substantially more advantageous (1T1R1T and 1T1R1T-VGA) than conventional 2T1R-RV in attainable retention (Fig. 6), thanks to the notably reduced BL resistance and improved drive current in the case of 1T1R1T (Table III). The conventional 2T1R-RV, in contrast, is only competitive in smaller (≤ 64 rows) arrays over 1T1D1R in terms of writability (Fig. 6) where the periphery area overhead becomes more prominent, which given its already oversized bitcell footprint (Fig. 4) does not necessarily lend itself to the desired LLC capacity in the *cross-node matching* context.

We further acknowledge that, the Δ-to-retention conversion above (and consequently the associated writability analysis based thereupon) is inevitably somewhat arbitrary as the required Δ in reality is dictated by the error rate in question [28] per Eq. (3)

$$\Delta = -\log\left(-\frac{\tau_0}{\tau_{ret}} \log\left(1 - \frac{F}{N_{bits}}\right)\right), \quad (3)$$

where $\tau_0$ is the reciprocal of the attempt frequency (~ 1 GHz), $\tau_{ret}$ the retention time, $F$ the number of failures and $N_{bits}$ the total number of bits in the memory macro consisting of multiple arrays; the ratio $F/N_{bits}$ hence defines the error rate. Clearly, the *actual* attainable retention in a memory macro needs to be refined in a context-aware, error rate-specific manner in future studies.

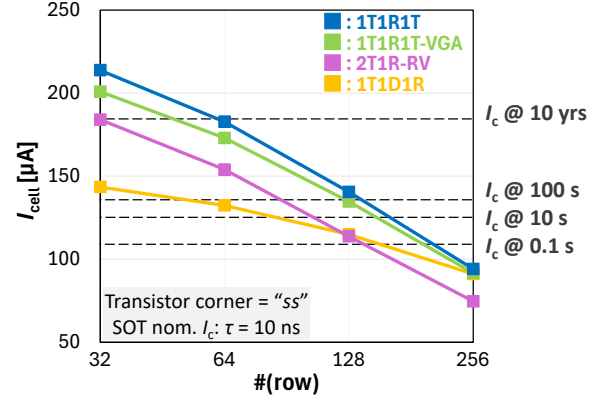

Fig. 6. Bitcell current ("*ss*" corner) for the different cell configurations in Table III versus the subarray size measured in #(rows). Also indicated are the nominal SOT switching current corresponding to different retention targets in Fig. 5 (at 10 ns write pulse width).

## V. Readability Assessment with Read Selectors

The shift to selector-based reading in SOT-MRAM bitcells is anything but self-evident insofar as the read voltage, speed and energy are all tied to the electrical characteristics of selectors. Therefore we have developed compact models (CMs) for the planar IGZO-FET and both symmetric tunnel- and asymmetric Schottky-diodes (Fig. 7) that are calibrated to experimental data or hardware-derived TCAD [30], for 1T1R1T and 1T1D1R, respectively. The 1T1R1T-VGA read performance is not evaluated for the lack of electrical data on VGA IGZO-FET at the time being [17]. Moreover, given the exploratory status of the BEOL-RS devices where even the standalone fabrication processes are undergoing multiple iterations, we assume it premature to identify fairly the "process corners" thereof, for which we will solely focus on their representative, typical characteristics in the read analysis.

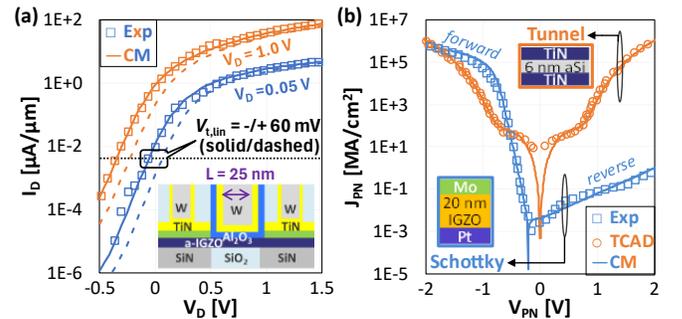

Fig. 7. CM calibration on (a) an experimentally measured planar IGZO-FET with a 25-nm channel length and a slightly negative $V_{t,lin}$ (i.e., $V_t$ at $V_D$ = 0.05 V), and on (b) a TCAD-simulated [30] TiN/aSi/TiN tunnel diode (with 6 nm amorphous Si or "aSi") and a measured Mo/IGZO/Pt Schottky diode (with 20 nm IGZO). Note that, in (a) the dashed, parallelly right-shifted curves depict a fictitious device with a positive $V_{t,lin}$, and in (b) the Schottky diode operates in forward regime at $V_{PN}$ < 0.

The bitcell readability is assessed in a 2-KB (128×128) subarray in SPICE [23] simulations using a precharge-and-discharge voltage sensing technique [31], where the MTJ parallel ("P") and antiparallel ("AP") resistance states are discriminated by the readline (RL) or sourceline (SL) discharging rate (Fig. 8). A minimum sensing margin (SM) of 100 mV is stipulated to compensate eventual noises, coupling and offset [32]. We canvass the parameter space in Table II and on top of that the read voltage ($V_{read}$), and extract key circuit metrics from read simulations (Table V). In view of the pathfinding nature of this work and the focus being bitcell exploration, we leave the *exact technology assumptions and layout/area of periphery circuitry for future work* once consensus is reached on bitcell design. The implications of eventual process variations in the periphery, e.g., in the sensing amplifiers are hence lumped in the minimum SM requirement [32] as a crude first-order estimation.

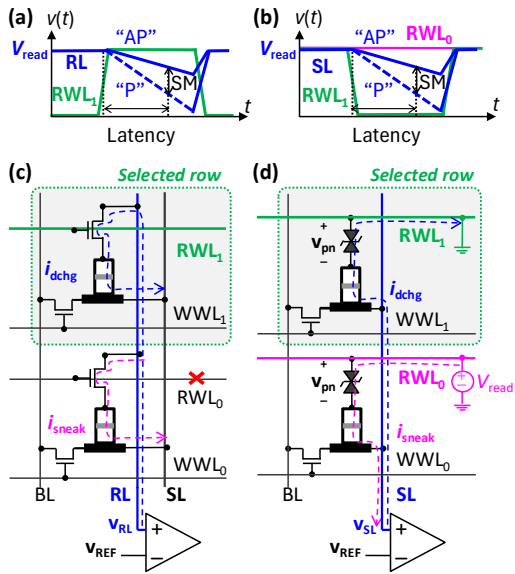

Fig. 8. (a)(b) Simplified timing diagram in voltage-based sensing, for (c) 1T1R1T [1] and (d) 1T1D1R arrays, respectively. SM is created during discharging of the selected bitcell by discharging current $i_{dchg}$. Note that the sneaky current $i_{sneak}$ (c) in 1T1R1T runs through transistor drain junction and carries the same sign as $i_{dchg}$; (d) in 1T1D1R $i_{sneak}$ opposes $i_{dchg}$ through unselected RWLs held at $V_{read}$, which is small thanks to the nonlinearity and/or rectifying effect of the diode.

TABLE V. READ DESIGN SPACE IN 2-KB (128×128) ARRAY

|  | 2T1R RV | 1T1R1T $V_{t,lin}$ = -60 mV (as-is) | 1T1R1T $V_{t,lin}$ = +60 mV (extrap.) | 1T1D1R Tunnel (sym.) | 1T1D1R Schottky (asym.) |
|---|---|---|---|---|---|
| RA [Ω·μm²] | 20 | 500 | 700 | 100 | 100 |
| TMR [%] | 86 | 200 | 110 | 200 | 150 |
| $R_{sel}$ [kΩ] | 3.4 | 186 | 338 | 361 | 197 |
| $TMR_{eff}$ [%] | 75 | 101 | 56 | 56 | 44 |
| $i_{dchg}$ [μA] | 25 | 0.83 | 0.66 | 4.0 | 4.5 |
| $|i_{dchg}/\Sigma i_{sneak}|$ [a] | 58 | 2.1 | 16 | 1.1 | 3.1 |
| $V_{read}$ [V] | 0.7 | 1.8 | 1.4 | 2.1 | 1.6 |
| SM [mV] | 107 | 102 | 108 | 99 | 104 |

[a]: assuming all unselected cells are in the low-resistance, "P" state.

The key findings in the readability exercises are as follows:

1) *IGZO-FET read challenges: low drive and junction leakage*

Given its amorphous oxide channel, the BEOL IGZO-FET in 1T1R1T understandably suffers from both low ON-state drive current (< 100 μA/μm) and high OFF-state leakage (> 700 nA/μm; Fig. 7(a)). The consequences are twofold:

- **The low drive current** in IGZO-FET leads to a huge equivalent resistance ($R_{sel}$; see Table V) that sits in series with the MTJ, hence obscuring the difference between the between "AP" and "P" states which dilutes the effective TMR ($TMR_{eff}$) of the bitcell. Consequently, the 1T1R1T requires both an increased nominal TMR but also around 30× MTJ RA (Table V) for the MTJ to manifest its own resistance (Fig. 9). The impact on RC delay will be addressed in subsection 3).

- **The OFF-state leakage** associated with the negative $V_{t,lin}$ adds to notable sneaky current ($i_{sneak}$; Fig. 8(c)) through the drain junction in unselected rows. Since the $i_{sneak}$ summed up from unselected rows is as visible to the column comparator as the discharging current ($i_{dchg}$), with high leakage it effectively swamps the $i_{dchg}$-to-$i_{sneak}$ ratio (Table V) and aggravates the available SM (Fig. 9).

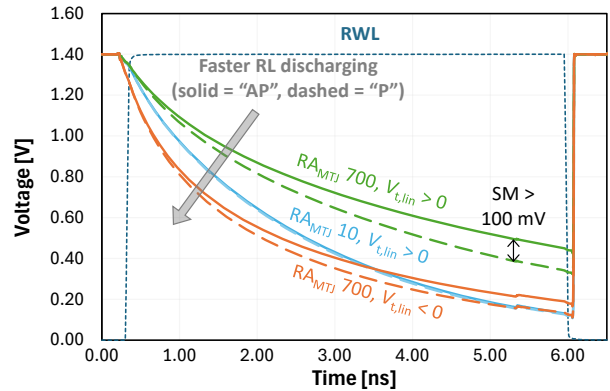

Fig. 9. SPICE [23]-simulated 1T1R1T RL discharging waveform during sensing, with three different MTJ RA and IGZO-FET $V_{t,lin}$ combinations (RA unit: Ω·μm²; $|V_{t,lin}|$ = 60 mV as in Table V). Decreasing either RA or $V_t$ increases RL discharging speed but undermines SM development, and only with a large RA and positive $V_{t,lin}$ is SM > 100 mV possible.

It is revealed that between these two detrimental limitations of the IGZO-FET it is the OFF-state leakage that is more critical to the readability challenge, which can be demonstrated by introducing a fictitious parallel positive $V_t$ shift to the IGZO-FET device model (dashes in Fig. 7(a)). Experimentally this is justified by the fact that the $V_t$ in IGZO-FET is largely controlled by the oxygen vacancy-induced conductivity of the channel and that proper annealing during device fabrication can effectively shift $V_t$ to positive at the expense of $I_{ON}$ [16]. Applying the positive $V_{t,lin}$ assumption to the 1T1R1T bitcell, one immediately notices the suppression of $i_{sneak}$ and the increase in $i_{dchg}$-to-$i_{sneak}$ ratio by 6×, which relaxes both the required TMR (150 %) and the $V_{read}$ (1.4 V) while still reaching the SM, even at increased $R_{sel}$ (Table V). Considering the latest development of IGZO-FET hardware with both low $I_{OFF}$ and high $I_{ON}$ [16][33] it is plausible that high(er) drive could be obtained in IGZO-FETs while still maintaining a positive $V_t$.

2) *Diode read challenge: imperfect rectification*

As with 1T1R1T, the 1T1D1R reading here uses voltage-based sensing (Fig. 8(d)) in order to get rid of DC current in the array and the consequent, undesired Joule dissipation. The key here however lies in proper inhibition of unselected rows since the common DC "half-bias" technique in current-based sensing

---

[1] The array read scheme for 2T1R is basically identical to 1T1R1T except that the RWL and WWL are merged in 2T1R (Fig. 2(a)).

[18][20] no longer applies. Instead, we resort to a "dynamic inhibition" holding the unselected RWLs at the precharge level $V_{read}$ (e.g., $RWL_0$ in Fig. 8(d)), such that at time zero only the selected cell sees $V_{read}$ between SL and $RWL_1$ whereas the unselected approximately zero. Granted, with SL being discharged $i_{sneak}$ does develop as SL voltage begins to fall under $RWL_0$ (Fig. 8(b)), but given the limited voltage drop between $RWL_0$ and SL during the development of SM it is anticipated that the nonlinear diode *I-V* characteristics (Fig. 7(b)) can still sufficiently suppress $i_{sneak}$ while maintaining $i_{dchg}$ that is biased close to $V_{read}$. This is particularly the case for the asymmetric Schottky diode, where $i_{dchg}$ and $i_{sneak}$ essentially flow in the forward and reverse bias regimes, respectively (Fig. 8(d)), that provides (partial) rectification (Fig. 7(b)).

The viability of the voltage sensing is attested in array simulations, where the asymmetric Schottky diode as expected provides a greater $i_{dchg}$-to-$i_{sneak}$ ratio with respect to the symmetric tunnel diode, at similar $i_{dchg}$ and even more relaxed $V_{read}$ (Fig. 10) – thanks to its rectifying effect. It is nevertheless observed that such rectification is far from being ideal compared to transistor-based selections as in 2T1R and 1T1R1T with a positive $V_{t,lin}$ (Table V), where the $i_{dchg}$-to-$i_{sneak}$ ratio easily reaches $\sim 10^1$, for which a greater precharge $V_{read}$ has to be introduced for 1T1D1R to boost the SM. Whereas this is an inherent challenge to the symmetric tunnel diode, for the Schottky diode the imperfect rectification due to excessive reverse current has been shown to be caused by perimeter leakage resulting from unintended etching damage during fabrication [20]. Clearly, continued refinement of the processing of IGZO-based Schottky diode would be instrumental to reducing the $V_{read}$ in 1T1D1R arrays.

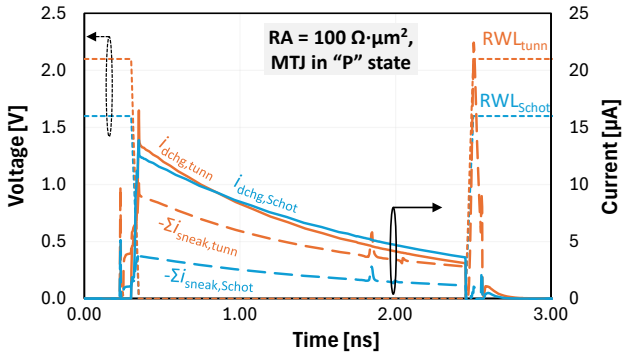

Fig. 10. SPICE [23]-simulated 1T1D1R discharging and sneaky current during sensing using the parameters in Table V; the MTJ is in "P" (low-resistance) state. The $i_{sneak}$ (as in Fig. 8(d)) is here plotted in its negative value in order to compare with the magnitude of $i_{dchg}$.

### 3) 1T1R1T vs 1T1D1R: tradeoff between speed and energy

Fig. 11 summarizes the array read power-performance of different bitcell configurations based on the design space in Table V. Evidently, the suboptimal $i_{dchg}$-to-$i_{sneak}$ ratio in the negative-$V_{t,lin}$ 1T1R1T not only hinders SM development but also directly adds to extra energy dissipation, accounting for > 70 % of the total read energy. Similar stories hold for the 1T1D1R bitcells where the lack of or imperfect rectification also penalizes energy consumption.

In the meantime, it is worth noticing that the 1T1D1R additionally sees a substantial increase in selector energy with ~ 5× higher read discharge current (Table V). This turns out to have to do with the highly nonlinear *I-V* characteristics of the diodes (Fig. 7(b)) needed to "self-select" the targeted cell, which incidentally brings in a relatively large $i_{dchg}$ to the selected cell that speeds up the sensing (~ 1.5 ns). Whereas for 1T1R1T the selection is explicitly executed by gate field effect of the IGZO-FET that then does not necessitate a very large $I_{ON}$ for SM development *per se* – if one allows it to occur in a "trickling" manner – which happens to perfectly accommodate its low-drive. Considering the extra RC delay already introduced by the large MTJ RA to salvage $TMR_{eff}$ (Table V), it is then unsurprising that the 1T1R1T reading overall appears rather time-consuming, as indicated by its (3 – 5) ns latency. Hence it is the difference in selection method (self-selection versus field effect) that crucially determines the tradeoff in speed and energy between 1T1D1R and 1T1R1T, with 1T1D1R tending to be more power-hungry and 1T1R1T more sluggish.

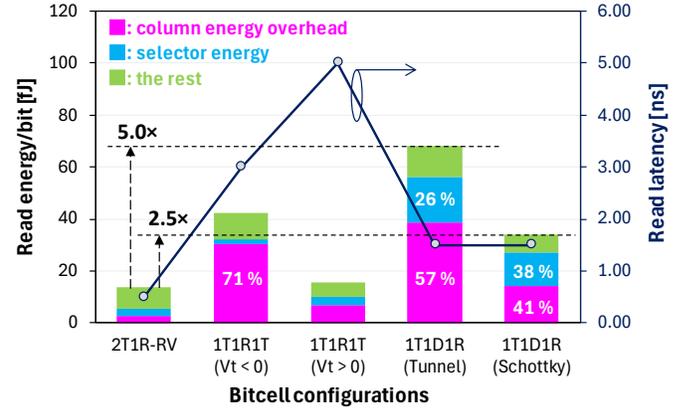

Fig. 11. SPICE [23]-simulated bitcell read power-performance in a 2-KB (128×128) array based on the design parameters outlined in Table V, with read energy breakdown in column overhead (including sneaky current), selector itself and the rest.

## VI. CONCLUSIONS

This paper explored the idea of *cross-node footprint-matched* SOT-MRAM as a cost-effective high-capacity heterogeneous last-level cache and exemplified its viability with extensive bitcell DTCO exercises at 7 nm node. We critically pointed out that the BEOL MTJ routing forms the top challenge to bitcell area miniaturization in conventional 2T1R SOT-MRAMs and demonstrated the read selectors (particularly diodes) as an effective, processing-friendly scaling enabler that unlocks sub-N3 SRAM bitcell footprint (0.02 μm$^2$). We have additionally confirmed up to 22 % write current boost in read selector-based arrays benefiting from both architectural and dimensional scaling of the bitcell, that promises to meet the SOT switching current in LLC-oriented magnetic layers with a retention target of (0.1 – 100) seconds. Last but not least, we have identified the read speed-versus-energy tradeoff between 1T1R1T (up to 5 ns latency) and 1T1D1R bitcells (up to 5× energy of 2T1R) and further highlighted the quest for high-drive, low-leakage FET selectors and rectification-improved diode selectors to achieve sub-ns and energy-efficient reading.

### ACKNOWLEDGMENT
The authors would like to thank D. Biswas for the inspiration.